\documentstyle[12pt]{article}

\begin{document}
\setcounter{page}{1}
\pagestyle{plain} \vspace{1cm}
\begin{center}
\Large{\bf Cosmological dynamics with modified induced gravity on the normal DGP branch}\\
\small \vspace{1cm} {\bf Kourosh
Nozari}\footnote{knozari@umz.ac.ir}\quad and
\quad{\bf Faeze Kiani}\footnote{fkiani@umz.ac.ir}\\
\vspace{0.5cm} {\it Department of Physics,
Faculty of Basic Sciences,\\
University of Mazandaran,\\
P. O. Box 47416-95447, Babolsar, IRAN}\\

\end{center}
\vspace{1.5cm}
\begin{abstract}
In this paper we investigate cosmological dynamics on the normal
branch of a DGP-inspired scenario within a phase space approach
where induced gravity is modified in the spirit of $f(R)$-theories.
We apply the dynamical system analysis to achieve the stable
solutions of the scenario in the normal DGP branch. Firstly, we
consider a general form of the modified induced gravity and we show
that there is a standard de Sitter point in phase space of the
model. Then we prove that this point is stable attractor only for
those $f(R)$ functions that account for late-time cosmic speed-up.
\end{abstract}
\vspace{1.5cm}

\newpage
\section{Introduction}
There are many astronomical evidences supporting the idea that our
universe is currently undergoing a speed-up expansion [1]. Several
approaches are proposed in order to explain the origin of this novel
phenomenon. These approaches can be classified in two main
categories: models based on the notion of \emph{dark energy} which
modify the matter sector of the gravitational field equations and
those models that modify the geometric part of the field equations
generally dubbed as \emph{dark geometry} in literature [2,3]. From a
relatively different viewpoint (but in the spirit of dark geometry
proposal), the braneworld model proposed by Dvali, Gabadadze and
Porrati (DGP) [4] explains the late-time cosmic speed-up phase in
its self-accelerating branch without recourse to dark energy [5].
However, existence of ghost instabilities in this branch of the
solutions makes its unfavorable in some senses [6]. Fortunately, it
has been revealed recently that the normal, ghost-free DGP branch
has the potential to explain late-time cosmic speed-up if we
incorporate possible modification of the induced gravity in the
spirit of $f(R)$-theories [7]. This extension can be considered as a
manifestation of the scalar-tensor gravity on the brane. Some
features of this extension are studied recently [8,9].

Within this streamline, in this paper we study the phase space of
the normal DGP cosmology where induced gravity is modified in the
spirit of $f(R)$-theories. We apply the dynamical system analysis to
achieve the stable solutions of the model. To achieve this goal, we
firstly consider a general form of the modified induced gravity. We
obtain fixed points via an autonomous dynamical system where the
stability of these points depends explicitly on the form of the
$f(R)$ function. There are also de Sitter phases, one of which is a
stable phase explaining the late-time cosmic speed-up. Secondly, in
order to determine the stability of critical points and for the sake
of clarification, we specify the form of $f(R)$ by adopting some
cosmologically viable models. The phase spaces of these models are
analyzed fully and the stability of critical points are studied with
details.

\section{DGP-inspired $f(R)$ gravity}

\subsection{The basic equations}
Modified gravity in the form of $f(R)$-theories are derived by
generalization of the Einstein-Hilbert action so that $R$ (the Ricci
scalar) is replaced by a generic function $f(R)$ in the action
\begin{equation}
S=\int d^{4}x \sqrt{-g}\Big(\frac{f(R)}{2\kappa^{2}}+{\bf
L_{m}}\Big)\,,
\end{equation}
where ${\bf L_{m}}$ is the matter Lagrangian and $\kappa^{2} = 8\pi
G$. Varying this action with respect to the metric gives
\begin{equation}
G_{\mu\nu}=\kappa^{2}T^{(tot)}_{\mu\nu}=\kappa^{2}
\Big(T^{(m)}_{\mu\nu}+T^{(
f)}_{\mu\nu}\Big)=\kappa^{2}\frac{\widetilde{T}^{(m)}_{\mu\nu}+\widetilde{T}^{
(f)}_{\mu\nu}}{f'}
\end{equation}
where $\widetilde{T}^{(m)}_{\mu\nu}=diag(\rho,-p,-p,-p)$ is the
stress-energy tensor for standard matter, which is assumed to be a
perfect fluid and by definition $f'\equiv\frac{df}{dR}$\,. Also
$\widetilde{T}^{(f)}_{\mu\nu}$ is the stress-energy tensor of the
\emph{curvature fluid} that is defined as follows
\begin{equation}
\widetilde{T}^{
(f)}_{\mu\nu}=\frac{1}{2}g_{\mu\nu}\Big[f(R)-Rf'\Big]+f'^{\,\,;\alpha\beta}
(g_{\alpha\mu}g_{\beta\nu}-g_{\alpha\beta}g_{\mu\nu})\,.
\end{equation}
By substituting a flat FRW metric into the field equations, one
achieves the analogue of the Friedmann equations as follows [10]
\begin{equation}
3f' H^{2}=\kappa^{2}\rho_{m}+\Big
[\frac{1}{2}\Big(f(R)-Rf'\Big)-3H\dot{f'}\Big]
\end{equation}
\begin{equation}
-2f'\dot{H}=\kappa^{2}\rho_{m}+\dot{R}^{2}f'''+(\ddot{R}-H\dot{R})f''\,,
\end{equation}
where a dot marks the differentiation with respect to the cosmic
time. In the next step, following [9] we suppose that the induced
gravity on the DGP brane is modified in the spirit of $f(R)$
gravity. The action of this DGP-inspired $f(R)$ gravity is given by
\begin{equation}
S=\frac{1}{2\kappa_{5}^{3}}\int d^{5}x\sqrt{-g}{\cal R}+\int
d^{4}x\sqrt{-q}\bigg(\frac{f(R)}{2\kappa^{2}}+ {\bf{L}}_{m}\bigg)\,,
\end{equation}
where $g_{AB}$ is the five dimensional bulk metric with Ricci scalar
${\cal R}$, while $q_{ab}$ is induced metric on the brane with
induced Ricci scalar $R$. The Friedmann equation in the \emph{normal
branch} of this scenario is written as [9]
\begin{equation}
3f'H^{2}=\kappa^{2}\Big(\rho_{m}+\rho^{(f)}\Big)-\frac{3H}{r_{c}}\,,
\end{equation}
where
$r_{c}=\frac{G^{(5)}}{G^{(4)}}=\frac{\kappa_{5}^{2}}{2\kappa^{2}}$\,
is the DGP crossover scale with dimension of $[length]$ and marks
the IR (infra-red) behavior of the DGP model. The Raychaudhuri's
equation is written as follows
\begin{equation}
\dot{H}\Big(1+\frac{1}{2Hr_{c}f'}\Big)=-\frac{\kappa^{2}\rho_{m}}{2f'}
-\frac{\dot{R}^{2}f'''+(\ddot{R}-H\dot{R})f''}{2f'}\,\,.
\end{equation}
To achieve this equation we have used the continuity equation for
$\rho^{(f)}$ as
\begin{equation}
\dot{\rho}^{(f)}+3H\bigg(\rho^{(f)}+p^{(f)}+\frac{\dot{R}f''}{r_{c}(f')^{2}}\bigg)
=\frac{\kappa^{2}\rho_{m}\dot{R}f''}{(f')^{2}}\,,
\end{equation}
where the energy density and pressure of the \emph{curvature fluid}
are defined as follows
\begin{equation}
\rho^{(f)}=\frac{1}{\kappa^{2}}\bigg(\frac{1}{2}\Big[f(R)-Rf'\Big]
-3H \dot{f'}\bigg)\,,
\end{equation}
\begin{equation}
p^{(f)}=\frac{1}{\kappa^{2}}\bigg({2H
\dot{f'}+\ddot{f'}-\frac{1}{2}\Big[ f(R)-R f'\Big]}\bigg)\,.
\end{equation}

After presentation of the required field equations, we analyze the
phase space of the model fully to explore cosmological dynamics of
this setup.

\subsection{ A dynamical system viewpoint}
The dynamical system approach is a convenient tool to describe
dynamics of cosmological models in phase space. In this way, we
rewrite equation (7) in a dimensionless form as
\begin{equation}
1=\frac{\rho_{m}}{3H^{2}f'}-\frac{1}{Hr_{c}f'}+
\frac{f(R)}{6H^{2}f'}-\frac{R}{6H^{2}}-\frac{\dot{f'}}{Hf'}\,.
\end{equation}
In the present study, we firstly consider a generic form of the
$f(R)$ function, so that one can define the dynamical variables
independent on the specific form of the $f(R)$ function as follows
(see for instance Ref. [10])
\begin{equation}
x_{1}=\frac{\rho_{m}}{3H^{2}f'}\,,\quad\
x_{2}=-\frac{1}{Hr_{c}f'}\,,\quad\ x_{3}=\frac{f}{6H^{2}f'}\,,\quad\
x_{4}=-\frac{R}{6H^{2}}\,,\quad\ x_{5}=-\frac{\dot{f'}}{Hf'}\,.
\end{equation}
Also we define the following quantities
\begin{equation}
m\equiv\frac{d \ln{f'}}{d \ln{R}}=\frac{Rf''}{f'}
\end{equation}
\begin{equation}
\quad\quad\ r\equiv-\frac{d \ln{f}}{d
\ln{R}}=-\frac{Rf'}{f}=\frac{x_{4}}{x_{3}}\,.
\end{equation}
We note that a constant value of $m$ leads to the models with
$f(R)=\xi_{1}+\xi_{2} R^{1+m}$ where the parameter $m$ shows the
deviation of the background dynamics from the standard model and
$\xi_{1}$ and $\xi_{2}$ are constants. However, in general the
parameter $m$ depends on $R$ and $R$ itself can be expressed in
terms of the ratio $r=\frac{x_{4}}{x_{3}}$ . This means that $m$ is
a function of $r$,\,that is, $m=m(r)$. Based on the new variables,
the Friedmann equation becomes a constraint equation so that we can
express one of these variables in terms of the others. Introducing a
new time variable $\tau=\ln{a}=N$\, and eliminating $x_{1}$\,(by
using the Friedmann constraint equation) we obtain the following
autonomous system
\begin{equation}
\frac{d x_{2}}{dN}=x_{2}(x_{5}+x_{4}+2)\,,\
\end{equation}
\begin{equation}
\quad\quad\quad\quad\,\,\, \frac{d
x_{3}}{dN}=-\frac{x_{4}x_{5}}{m}+x_{3}(2x_{4}+x_{5}+4)\,,\
\end{equation}
\begin{equation}
\quad\,\,\, \frac{d
x_{4}}{dN}=\frac{x_{4}x_{5}}{m}+x_{4}(2x_{4}+4)\,,\
\end{equation}
\begin{equation}
\frac{d
x_{5}}{dN}=(x_{2}+x_{5})(x_{5}+x_{4})+1-3x_{3}-5x_{4}-2x_{2}\,,\
\end{equation}
and
\begin{equation}
x_{1}\equiv\Omega_{m}=1-x_{2}-x_{3}-x_{4}-x_{5}\,.\
\end{equation}

The deceleration parameter which is defined as
$q=-1-\frac{\dot{H}}{H^{2}}$\,,\ now can be expressed as
\begin{equation}
q=1+x_{4}\,,\
\end{equation}
and the effective equation of state parameter of the system is
defined by
\begin{equation}
\omega_{eff}=-1-\frac{2\dot{H}}{3H^{2}}\,\,.
\end{equation}

\begin{figure}[htp]
\begin{center}\includegraphics{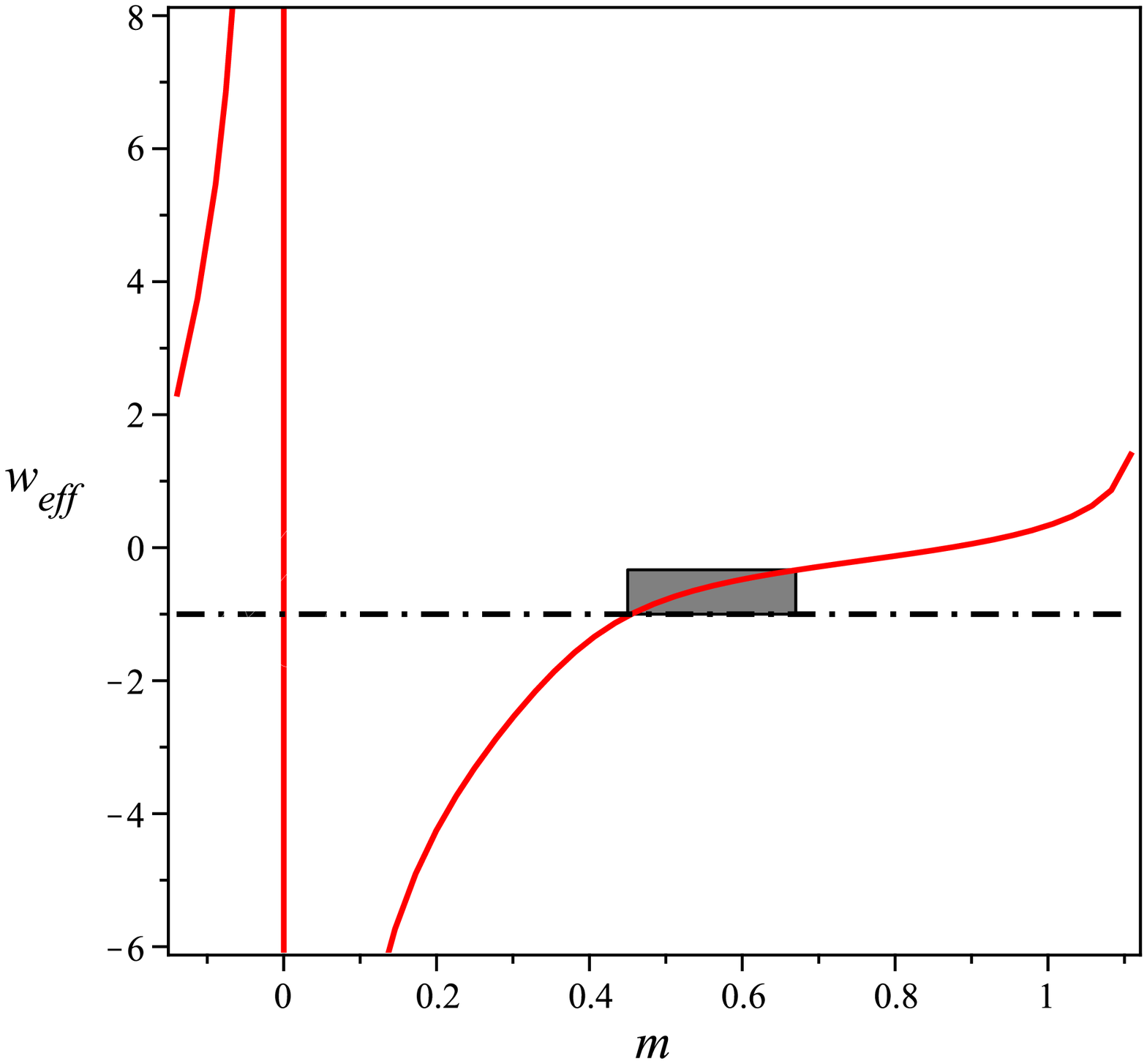} \vspace{2.3cm}\includegraphics{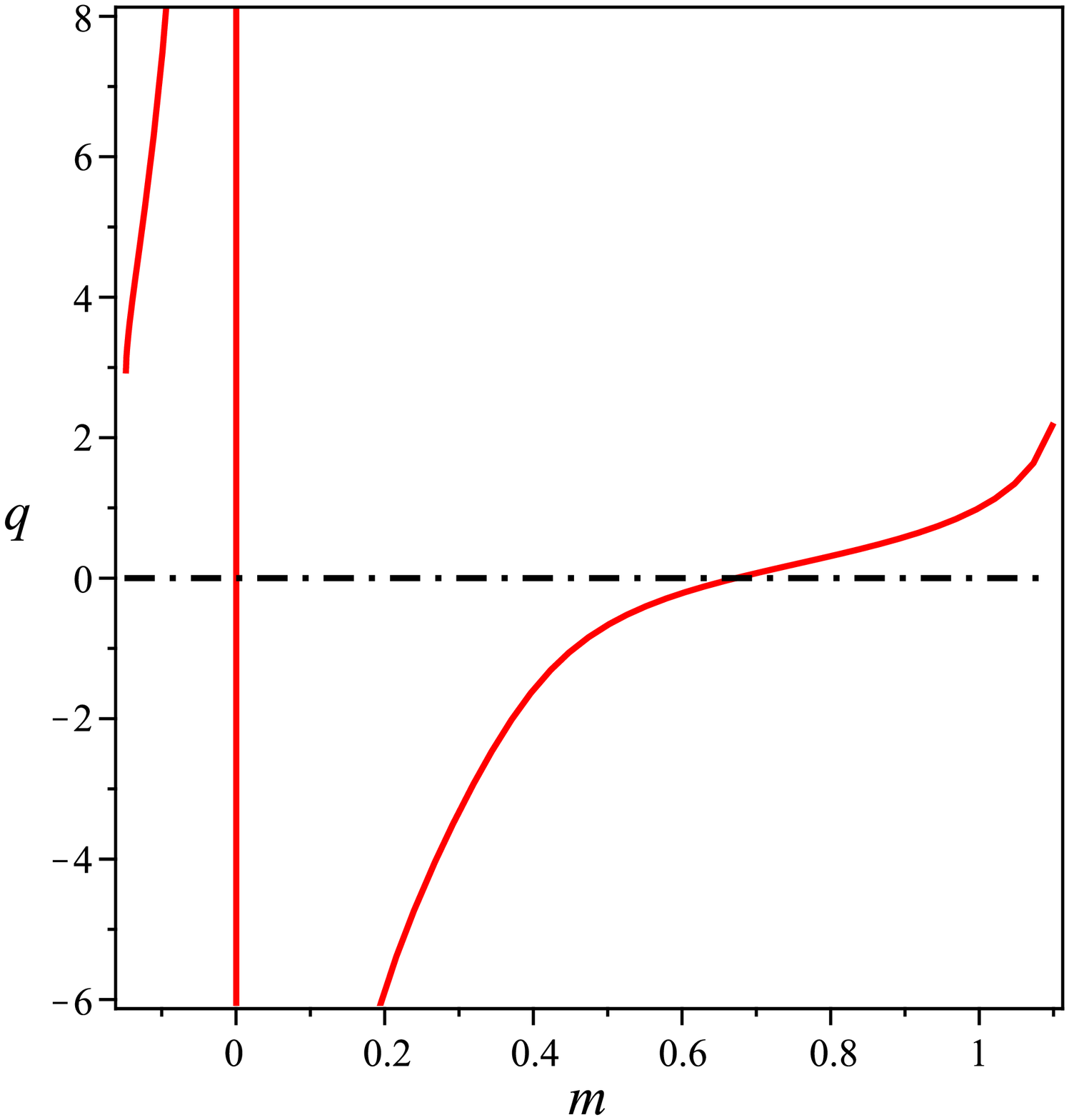} \vspace{3.4cm}
\end{center}
 \caption{\small {The effective equation of state parameter (left panel) and the deceleration
 parameter (right panel) of the critical point ${\cal E}$ versus $m$. The
 shaded region denotes a non-phantom acceleration era which occurs in the parametric space with $0.45<m<0.67$\,.
 A slightly phantom-like behavior exists for $m<0.45$ and when
 $m\rightarrow 0$\,, it reaches to a strong phantom-like phase.
 For negative values of $m$ the EoS parameter is stiff-like.  }}
\end{figure}

\subsection{Critical points and their stability}

The critical points of the scenario and some of their properties are
listed in table $1$. In this table, $\Gamma$ is defined as
$$\Gamma\equiv\frac{1}{2}\,{\frac {4\,{m}^{2}-9\,m+2\pm\sqrt
{-160\,{m}^{4}+272\,{m}^{3}-111\,
{m}^{2}+4\,m+4}}{2\,{m}^{2}-3\,m+1}}\,.$$ We consider only the plus
sign of this equation in our forthcoming arguments. The minus sign
does not create suitable cosmological behavior since it leads to
$w_{eff}<-10$ or $w_{eff}>0.7$ for point $\cal{E}$.

In table $1$, the critical points $\cal{A}$\,, $\cal{B}$ and
$\cal{C}$ are independent of the form of $f(R)$. Nevertheless, the
stability of these points depends on the form of $f(R)$ explicitly.
The critical curve ${\cal D}$ exists just for $f(R)$ models with
$m(r=-\frac{1}{2})=\frac{1}{2}$ (for instance, in models of the form
$f(R)=R+\gamma R^{-n}$\, that $m$ is defined as
$m(r)=-\frac{n(1+r)}{r}$,\, the critical curve ${\cal D}$ exists
just for $n=\frac{1}{2}$). The value of the effective equation of
state parameter corresponding to this critical curve depends on the
coordinate $x_{2}^{*}$\,. Then, different intervals on this curve
describe different era of the universe evolution (for example, a
formal de Sitter-type era occurs for curve ${\cal D}$ at
$x^{*}_{2}=-\frac{1}{4}$). Since the phase space behavior of the
point ${\cal E}$ (with $m(r)=1+r$) depends on the form of $f(R)$ via
$m$, we treat it separately in section $3$. In which follows, we
classify the important subclasses in order to see their dynamical
behaviors.\\

\newpage

{\bf a) The radiation dominated era }\\
\\
The points ${\cal A}$ and ${\cal B}$ demonstrate effectively the
radiation dominated epoch of the universe
($\omega_{eff}=\frac{1}{3}$ with $a(t)\propto t^{\frac{1}{2}}$).
This phase can be realized also by the curve ${\cal D}$ for
$x^{*}_{2}=\frac{5}{4}$\,. The stability of these points is
investigated in which follows.\\

{\bf b) The matter dominated era }\\
\\
The matter era ($\omega_{eff}=0$) could be existed for models with
$m(r=-\frac{1}{2})=\frac{1}{2}$ by the localized point
$x^{*}_{2}=\frac{7}{8}$ on the curve ${\cal D}$\,. Note that this
matter era is properly described by a cosmic expansion with scale
factor $a(t)=a_{0}(t-t_{0})^{\frac{2}{3}}$. This era also can be
realized in localized point $\cal{E}$ with $m(r=-0.13)=0.87$\,.\\

{\bf c) The de Sitter era }\\
\\
The de Sitter phase ($\omega_{eff}=-1$) in the normal branch of this
DGP-inspired $f(R)$ model is realized by the curve ${\cal C}$ of
critical points. It is important to point out that the mentioned de
Sitter solution is the standard de Sitter phase just for
$x^{*}_{2}=2$\,, since in this localized point the matter density
parameter vanishes ($\Omega_{m}= 0$). Also this phase can be
realized from the curve of the non-localized points ${\cal D}$ with
$x^{*}_{2}=-\frac{1}{4}$\,\,(this de Sitter point can be regarded as
the special case of the curve ${\cal C}$ in the localized point
$x^{*}_{2}=-\frac{1}{4}$). But one can see that this point gives no
standard de Sitter era since its $\Omega_{m}$ is non-vanishing (in
this case $\Omega_{m}=0$ occurs at
$x^{*}_{2}=\frac{1}{5}$).\\

{\bf d) Transition from $q>0$ to $q<0$ }\\
\\
The critical point ${\cal E}$ depends explicitly on the form of
$f(R)$ via $m$ and this is the case also for stability of this
point. The point ${\cal E}$ describes a phase transition of the
universe from deceleration to the acceleration era at
$\omega_{eff}=-\frac{1}{3}$ and $q=0$\, in this model for
$m(r=-0.33)=0.67$\,. Also the mentioned feature for $f(R)$ models
with $m(r=-\frac{1}{2})=\frac{1}{2}$ (corresponding to the curve
${\cal D}$) occurs at the fixed point $x^{*}_{2}=\frac{1}{2}$\,. So,
one of the important feature of this model is that it clearly
realizes the late-time acceleration of the universe in its
\emph{normal branch} for $0< m\leq 0.67$\,. In figure $1$ the
localized point ${\cal E}$ represents the deceleration phase for
$m>0.67$ and a non-phantom accelerating phase for $0.45\leq m\leq
0.67$. For $m<0.45$\,, since $q<-1$\,, the model realizes an
\emph{effective phantom phase} with possibility of future big rip
singularity which is characteristics of a non-canonical (phantom)
field dominated universe. Similarly, the effective phantom behavior
for the curve ${\cal D}$ occurs at $x^{*}_{2}< -\frac{1}{4}$\,.\\
\begin{table}
\begin{center}
\caption{Location, effective EoS and deceleration parameter of the
critical points for the normal branch of a general DGP-inspired
$f(R)$ scenario.  The fixed point $\cal{D}$ exists only for those
$f(R)$ models that $m(r=-\frac{1}{2})=\frac{1}{2}$. } \vspace{0.5
cm}
\begin{tabular}{|c| c c c  c|c |c |c|}
  \hline
  \hline point &$x_{2}$ & $x_{3}$& $x_{4}$ & $x_{5}$&$r$   &$q$&$w_{eff}$ \\
  \hline
      ${\cal A}$ &$0$ &$\frac{17}{3}$&$0$ &$-4$&$0$ &$1$& $\frac{1}{3}$ \\
     \hline
      ${\cal B}$ &$\frac{5}{4}$ &$0$ & $0$ &  $-2$& indefinite  &$1$& $\frac{1}{3}$\\
     \hline
     ${\cal C}$ &$x^{*}_{2}$&$ \frac{1}{3}(11-4x^{*}_{2})$&$-2$&$0$&$\frac{6}{4x^{*}_{2}-11}$&
      $-1$ & $-1$\\
      \hline
      ${\cal
      D}$&$x^{*}_{2}$&$-\frac{2}{3}(4x^{*}_{2}-5)$&$\frac{1}{3}(4x^{*}_{2}-5)$&$-\frac{1}{3}(1+4x^{*}_{2})$&$-\frac{1}{2}$
      &$\frac{2}{3}(2x^{*}_{2}-1)$& $\frac{1}{9}\Big(8x^{*}_{2}-7\Big)$\\
      \hline
      $ {\cal E}$ &$0$ & $\frac{\Gamma+4m}{2m(1-m)}$ &
      $-\frac{\Gamma+4m}{2m}$& $\Gamma$&$m-1$& $-(1+\frac{\Gamma}{2m})$ & $-\Big(1+\frac{\Gamma}{3m}\Big)$\\
      \hline
      \hline
\end{tabular}
\end{center}
\end{table}

In the next step we determine the stability of the critical points
under small perturbations. The stability of these points is
determined by the eigenvalues of the Jacobian matrix. For a general
$f(R)$ term on the brane, stability of the critical points depends
on the form of $f(R)$ (or equivalently on the parameter $m$). It is
obvious that in general $m=m(r)$ is not a constant; it is a function
of other variables so that one can expand this function of curvature
about any of the fixed points. The results of our investigation for
stability of critical points mentioned in table $1$, are summarized
as follows:\\

$\bullet$ \, Point ${\cal A}$\\

As has been mentioned, this point is a radiation dominated era. In
this case the eigenvalues are
\begin{equation}
-2\,,\,\,\frac{4(m-1)}{m}\,,\,\,-4\pm\,i\,.
\end{equation}
Note that around this point, $m\equiv m_{\cal{A}}=m(r=0)$\,. Hence
this point is a spiral attractor if $0<m(r=0)<1$\,, otherwise it is
a saddle point. The corresponding 2D phase space for arbitrary
$m=m(r)$ is shown in figure $2$ (left panel). Note that the Jacobian
matrix in this point has no dependence on the $m'(r)$\, since this
point lies around $r=0$. Here a prime denotes derivative with the
respect to $r$\,.\\

$\bullet$ \, Point ${\cal B}$\\

This point is also a radiation dominated era in which the
eigenvalues are as follows
$$\lambda_{1,2}=-\frac{1}{8}(11\pm \sqrt{199}\,i)$$
\begin{equation}
\lambda_{3,4}={\frac{3\,m^{2}-m+m'r(1+r)}{m^{2}}\pm\frac{\sqrt
{(m^{2}-m)^{2}+2\,m'r{m}^{2
}(1-r)-2\,mm'r(1+r)+{m'}^{2}(r+{r}^{2})^{2}}}{m^{2}}}\,,
\end{equation}

\begin{figure}[htp]
\begin{center}\includegraphics{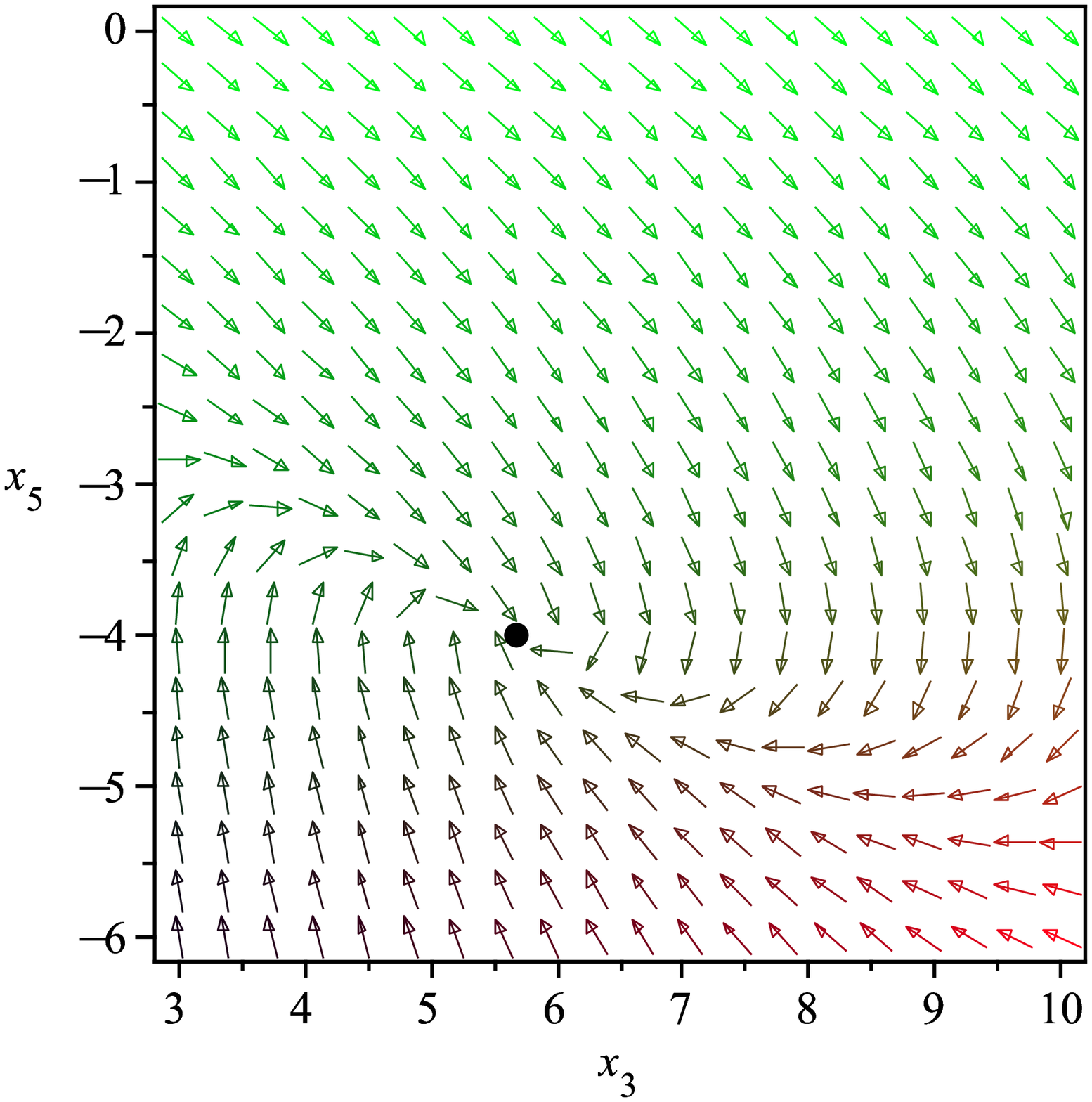}
\vspace{3cm}\includegraphics{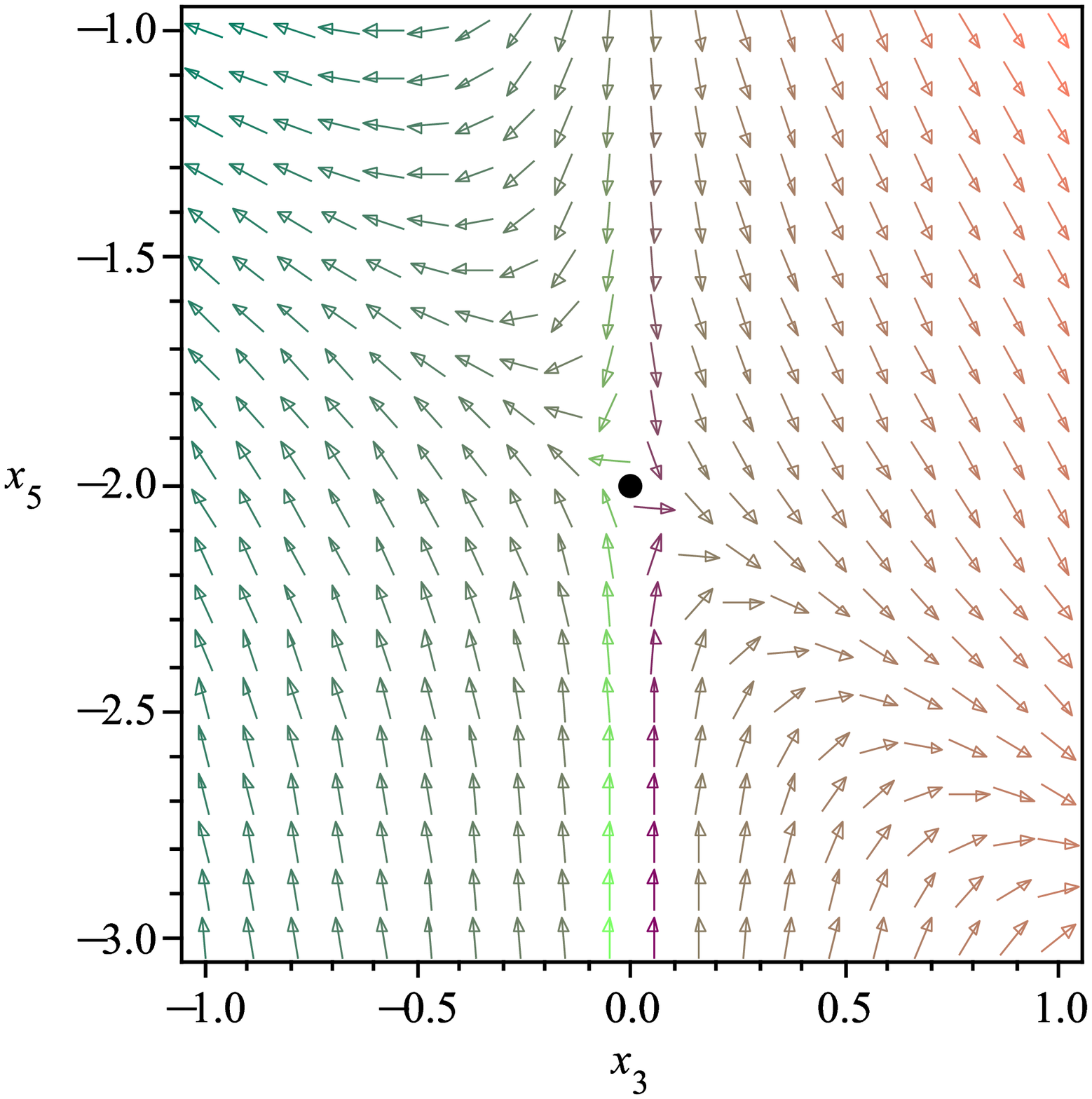} \vspace{2cm}
\end{center}
 \caption{\small {The phase subspace $x_{3}$ - $x_{5}$ of our setup at $x_{2}=x_{4}=0$ (left panel)
 and $x_{2}=\frac{5}{4}$ and $x_{4}=0$ (right panel)\,.
 The critical point shown in the left panel is point ${\cal A}$ and in the right panel it is point ${\cal B}$. Point ${\cal A}$ in the
 mentioned subspace is a spiral attractor and point ${\cal B}$ is a saddle point. Note that in the 4D phase space
these are either spiral attractor or saddle point depending on the
form of $f(R)$\,.}}
\end{figure}

where the parameter $m$ should be expanded around the point ${\cal
B}$ (that is, $m=m_{\cal{B}}$). Here the stability issue depends
also on $m$ and $m'$, so that this point can be either a saddle or a
spiral attractor. The corresponding 2D phase space for arbitrary
$m=m(r)$ is shown in figure $2$ (right panel). An important point
should be emphasized here: by setting $x_{4}=0$ in plotting figures
$2$, their dependence on $m(r)$ is wasted. However, the situation is
different if one plots the 3D subspace $x_{3}$ - $x_{4}$ - $x_{5}$
for these points. In this case, one should determine the form of the
function $m(r=\frac{x_{4}}{x_{3}})$. For a constant $m$, the
eigenvalues reduce to
\Big(2\,,\,$-\frac{1}{8}(11\pm\sqrt{199}\,i)$\,,\,$\frac{2(2m-1)}{m}$\,\Big)
which indicates that the point ${\cal B}$ is a saddle point for constant values of $m$.\\

$\bullet$ \, Curve ${\cal C}$\\

Generally, if a nonlinear system has a critical curve, the Jacobian
matrix of the linearized system at a critical point on the line has
a zero eigenvalue with an associated eigenvector tangent to the
critical curve at the chosen point. When dynamical variables are not
independent, some eigenvalues of the Jacobian matrix are zero. In
this case, the phase space of the nonlinear system reduces to a
lower dimensional phase space. The stability of an specific critical
point on the curve can be determined by the nonzero eigenvalues,
because near this critical point there is essentially no dynamics
along the critical curve (i.e., along the direction of the
eigenvector associated with the zero eigenvalue). So, the dynamics
near this critical point may be viewed in a reduced phase space
obtained by suppressing the zero eigenvalue direction. On the other
hand, such curves are actually \textit{normally hyperbolic} [11,12].
We consider a point on the curve ${\cal C}$ with coordinates  $(2,\,
1,\, -2,\, 0)$. This point is a standard de Sitter phase. The
eigenvalues corresponding to this point are as follows
\begin{equation}
0,\,\,\,\,\,  \frac{1}{3}\Big(\chi-\frac{17}{\chi}-4\Big),\,\,\,\,\,
-\frac{1}{6}\Big(\chi-\frac{17}{\chi}+8\Big)\pm
i\frac{\sqrt{3}}{6}\Big(\chi+\frac{17}{\chi}\Big)\,,
\end{equation}
where $\chi$ is defined as
\begin{equation}
\chi={\frac { \Big[\left( -460\,m+54+3\,\sqrt {3}\sqrt
{8019\,{m}^{2}- 1840\,m+108} \right)
{m}^{2}\Big]^{\frac{1}{3}}}{m}}.
\end{equation}

\begin{figure}[htp]
\begin{center}\includegraphics{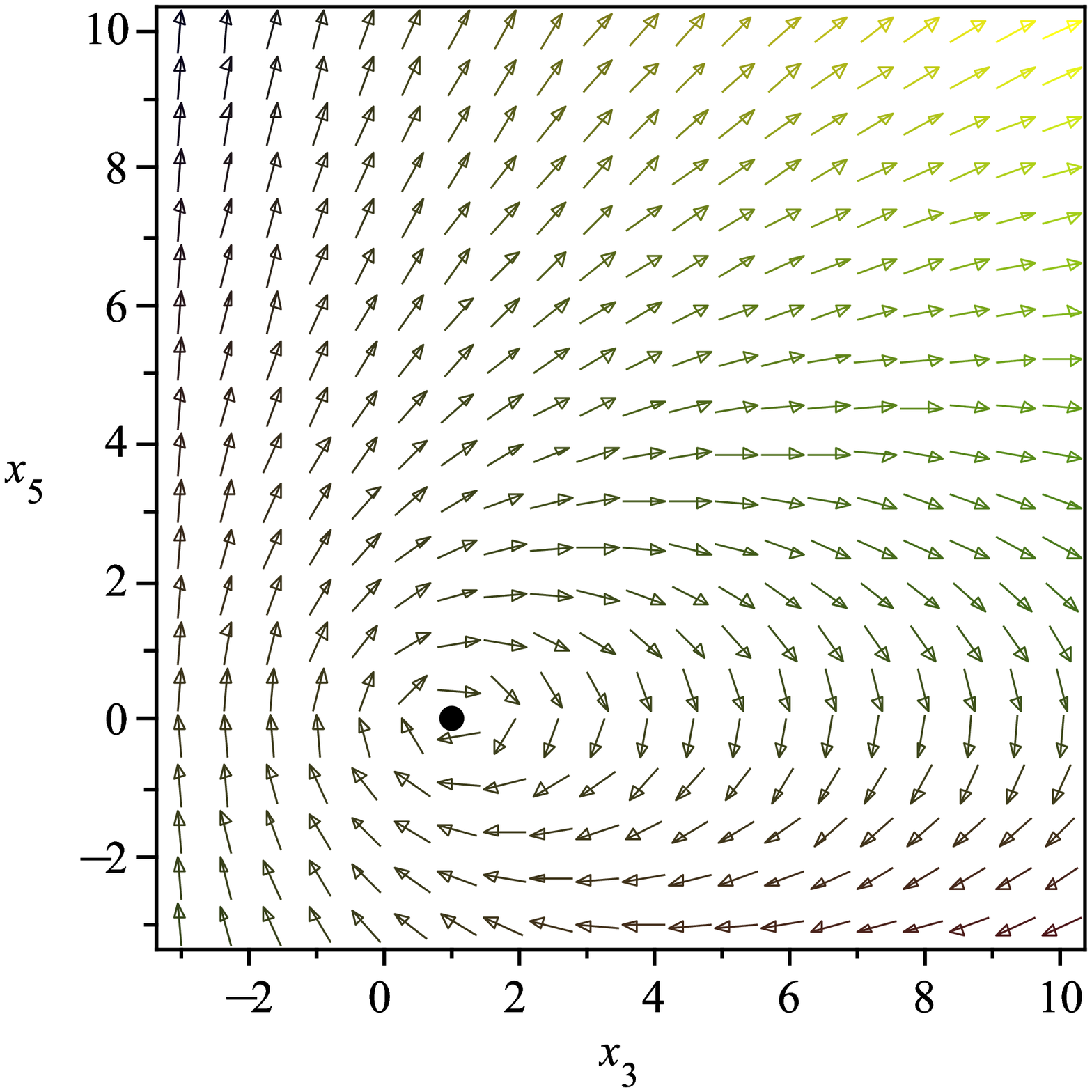} \vspace{3cm}\includegraphics{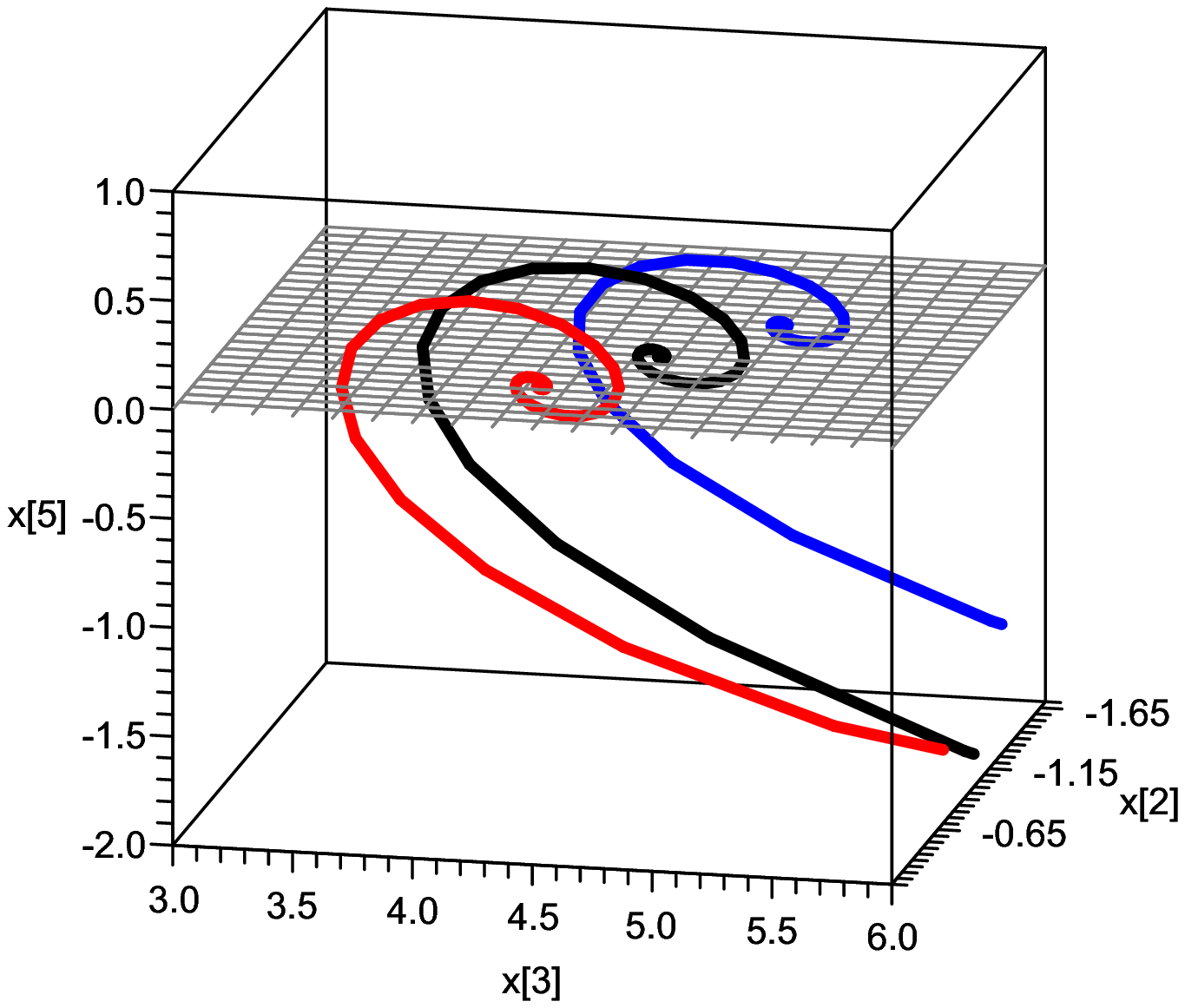} \vspace{2cm}
\end{center}
 \caption{\small { 2D phase space $x_{3}$ - $x_{5}$ for $x_{2}=2$ and $x_{4}=-2$ (left panel)
 and the 3D phase space $x_{2}$ - $x_{3}$ - $x_{5}$ for $x_{4}=-2$ (right panel).
 The critical point which is shown in the left panel is a center and is related to the standard
 de Sitter phase. The critical points
shown in the right panel which lie on the line
$x_{3}=\frac{1}{3}(11-4x_{2})$ are spiral attractors. These figures
are plotted for $m(r=-2)=\frac{1}{2}$. }}
\end{figure}
The parameter $m$ should be expanded around the standard point of
${\cal C}$ as defined previously with $m_{\cal{C}}=m(r=-2)$. Here
the stability issue depends only on $m$. It is a stable spiral in
the subspace of the last two eigenvalues when $0<m$. On the other
hand, the second eigenvalue is negative for $m<0$ and $0.09\leq m$.
So the standard de Sitter point in the $4$D phase space is a spiral
attractor if
\begin{equation}
m\geq0.09\,.
\end{equation}
This point for other values of $m$ is a saddle point. Figure 3 shows
the $2$D and $3$D phase spaces of the critical curve ${\cal C}$. We
note that in the $2$D subspace, the de Sitter curve is reduced to a
de Sitter point (as figure $3$ shows, this point is a center).
Therefore, the \emph{center manifold theory} is required to
investigate its stability [13,14]. In $3$D subspace (right panel),
the non-localized points in the $x_{2}$ - $x_{3}$ plane lie on the
line $x_{3}=\frac{1}{3}(11-4x_{2})$. Figure $3$ is plotted for
$m=\frac{1}{2}$ which satisfies (27), therefore this curve is a
spiral de Sitter attractor.\\

\section{Analytical results for some specific models }

As we have mentioned previously, $m$ is a function of $r$, that is,
$m=m(r)$. Since the $\Lambda$CDM model defined with
$f(R)=R-2\Lambda$, corresponds to $m=0$, we can say that the
quantity $m$ characterizes the deviation of the background dynamics
from the standard $\Lambda$CDM model. Now we consider two specific
model of $f(R)$ in order to obtain more obvious results. We also
focus on the cosmological viability of these models. A
cosmologically viable scenario contains an early time radiation
dominated era followed by a matter dominated era that reaches to a
standard de Sitter phase which is a stable attractor. We focus here
only on the last two stages: matter domination and then a stable de
Sitter attractor. At the first stage, the existence of a matter
dominated era (\,$w_{eff}=0$\,) constrains our DGP-inspired $f(R)$
model only to those $f(R)$ functions that
$m(r=-\frac{1}{2})=\frac{1}{2}$ and $m(r=-0.13)=0.87$ (see table 1).
In the first case, there exist two de Sitter phases: the localized
point $x_{2}^{*}=-\frac{1}{4}$ on the curve ${\cal D}$ which is a
formal de Sitter point, and also the de Sitter curve ${\cal C}$
which realizes a standard de Sitter point at the localized point
$x^{*}_{2}=2$. The second case is associated just to the de Sitter
curve ${\cal C}$. A correct connection between the unstable matter
dominated era and the stable, standard de Sitter era is necessary
condition for cosmological viability of a scenario.
This connection can be investigated in the $m-r$ plane.\\

\textbf{ A)\,\, $f(R)=R+\gamma R^{-n}$}\\
\\
For this model, the parameter $m$ takes the following form
\begin{equation}
m(r)=\frac{-n(1+r)}{r}\,,
\end{equation}
which is independent on $\gamma$\,. On the other hand, the point
${\cal E}$ of table $1$ is characterized by the following relation
\begin{equation}
m(r)=r+1\,.
\end{equation}
Equations (28) and (29) give two solutions $m_{1}=0$ and
$m_{2}=1-n$\, for $m$. So, for the mentioned $f(R)$ function, point
${\cal E}$ is characterized by the following relation
\begin{equation}
{\cal E}_{(1-n)}:\quad \Bigg(0,\,
\frac{\Gamma^{(1-n)}+4(1-n)}{2n(1-n)}\,,\
-\frac{\Gamma^{(1-n)}+4(1-n)}{2(1-n)}\,,\
\Gamma^{(1-n)}\Bigg)\,,\,\,\,\,\,\,
\omega_{eff}=-1-\frac{\Gamma^{(1-n)}}{3(1-n)}\,,
\end{equation}
where $\Gamma^{(1-n)}\equiv\Gamma|_{m=1-n}$.

Note that the critical point ${\cal E}$ for $m=0$ (that is, ${\cal
E}_{(0)}$) is indefinite and therefore we exclude it from our
considerations. Now we investigate the stability of the critical
point ${\cal E}_{(1-n)}$. Since the EoS parameter corresponding to
this point varies with $m$, in order to determine the stability of
this point, one has to fix the value of $m$. \\
In figure $4$ (left panel), we have shown the behavior of the
parameter $m$ as a function of $r$ for a special $f(R)$ model given
as $f(R)=R+\gamma R^{n}$\, with \,$n=0.13$\,. As this figure shows,
this model contains a connection between the matter era and the
standard de Sitter era which is located at $m=-0.065$. However, this
model is not cosmologically viable since it reaches an unstable
standard de Sitter era. In figure $4$ (right panel), we plotted the
curve $m(r)$ for $f(R)=R+\gamma R^{n}$ with $n=\frac{1}{2}$. The
matter dominated era evolves to the standard de Sitter era at
$m=-0.25$\, which indicates that the standard de Sitter era is
unstable (see Eq. (27)). So, this model is not cosmologically viable
too. Note that in the right panel of figure $4$, the point $A$ lies
also on the dashed line which is corresponding to the critical point
${\cal E}$. This feature indicates that this model contains the
critical point ${\cal E}$ with $m(r=-\frac{1}{2})=\frac{1}{2}$ which
is a non-phantom acceleration phase (compare this case with figure
$1$ that $m=\frac{1}{2}$ lies in the shaded region). \\
\\

\begin{figure}[htp]
\begin{center}\includegraphics{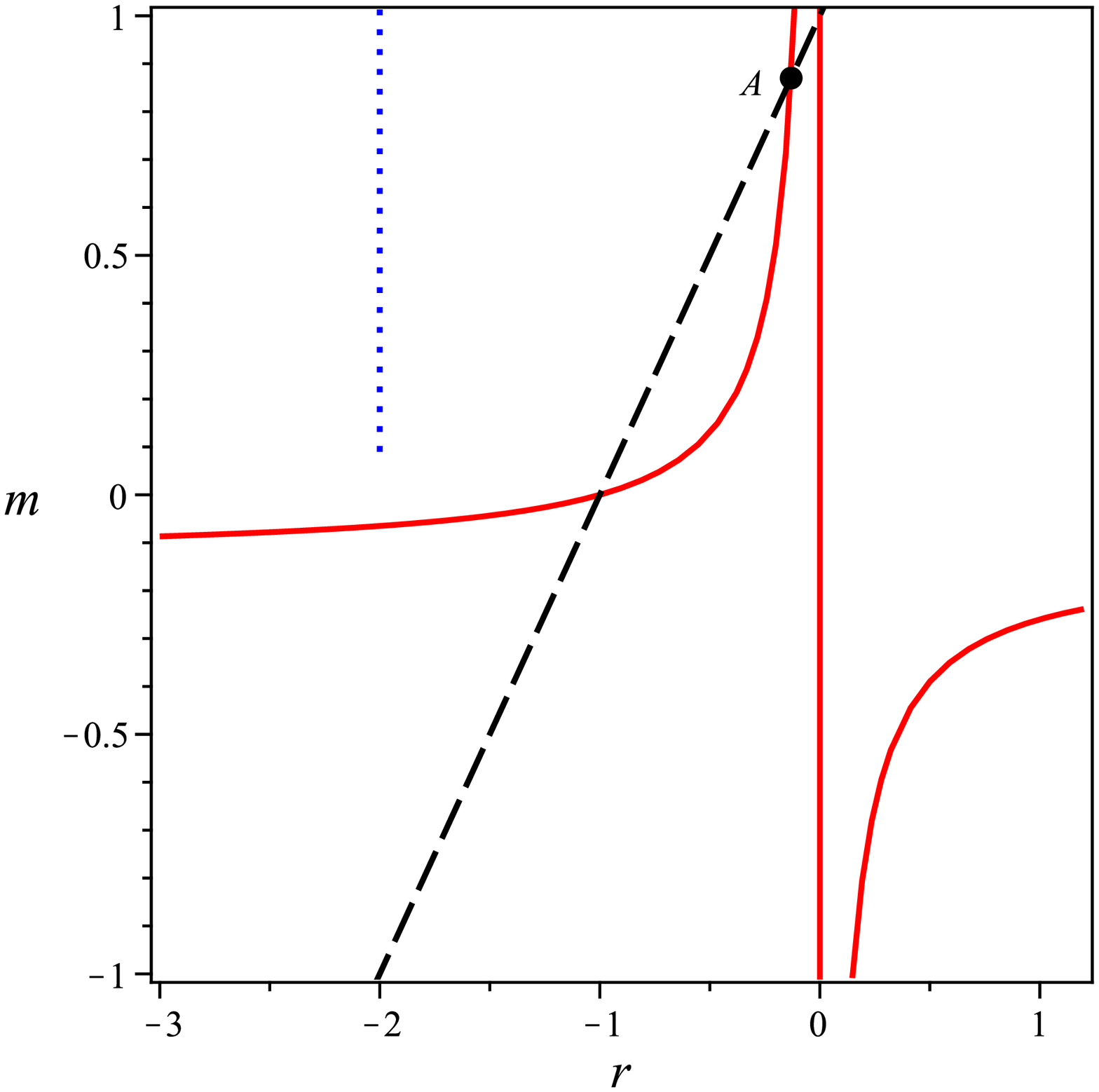}
\vspace{3cm}\includegraphics{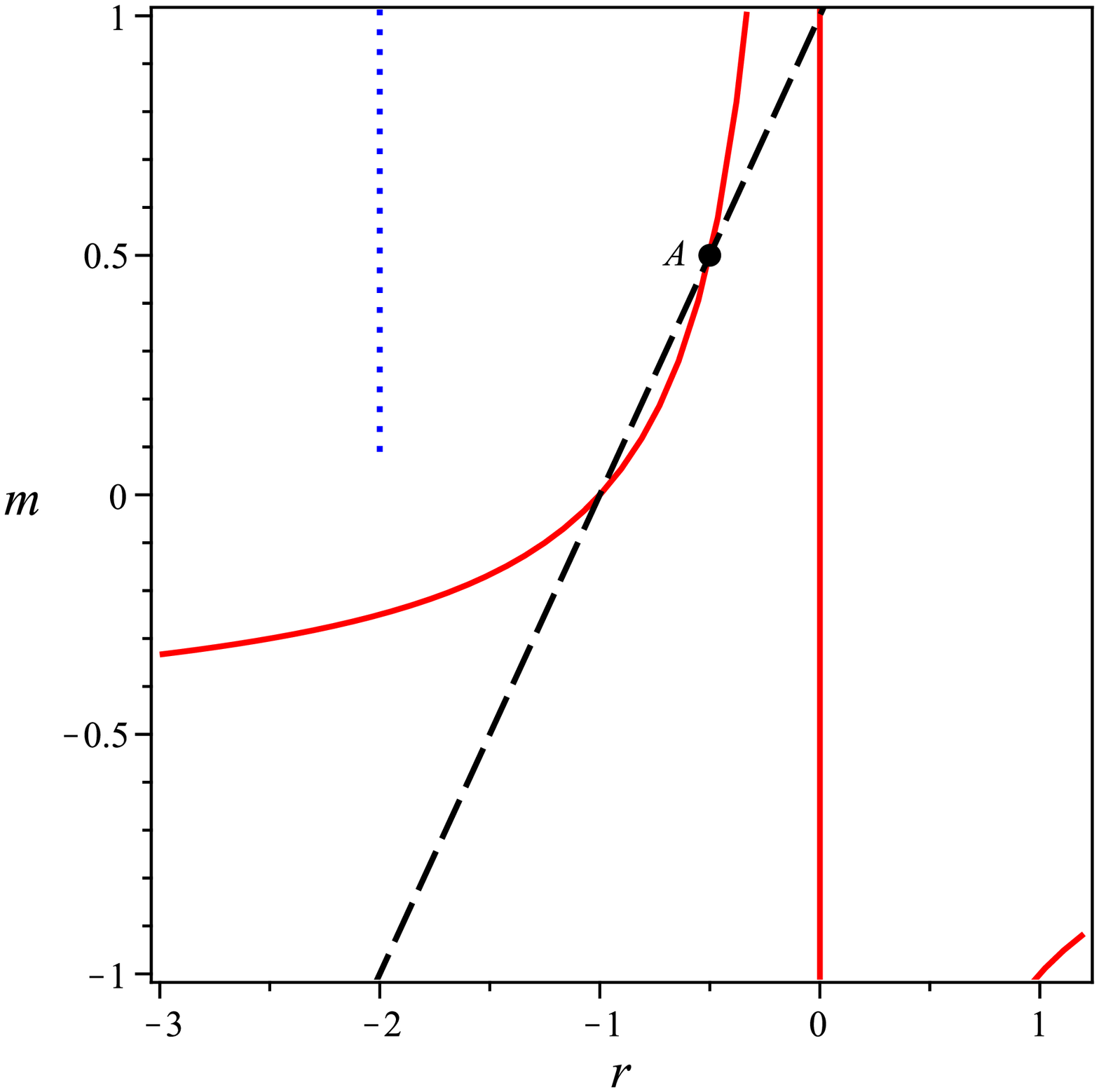} \vspace{2cm}
\end{center}
\caption{\small {The solid curves illustrate the diagram $m-r$ of
$f(R)=R+\gamma R^{-n}$ with $n=0.13$ (left panel) and with
$n=\frac{1}{2}$ (right panel). The dashed-line is corresponding to
the point ${\cal E}$ of table 1. The intersection point $A$ shows a
matter domination phase. The standard de Sitter phase is
corresponding to the line $r=-2$ and dotted (blue) line indicates
the line of stability.}}
\end{figure}

\textbf{ B)\,\, $f(R)=R^{^n} \exp(\frac{\eta}{R})$}\\

In this model the parameter $m$ is defined as
\begin{equation}
m(r)=-\frac{n+r(2+r)}{r}\,,
\end{equation}
which is independent on $\eta$\,. As has been pointed out
previously, the matter dominated era can be achieved from critical
point ${\cal E}$ in which the relative parameter is given by
equation (29). Equating (29) and (31), we obtain two solutions for
$m$ as $m=\frac{1\pm\sqrt{9-8n}}{4}$\,.
\begin{figure}[htp]
\begin{center}\includegraphics{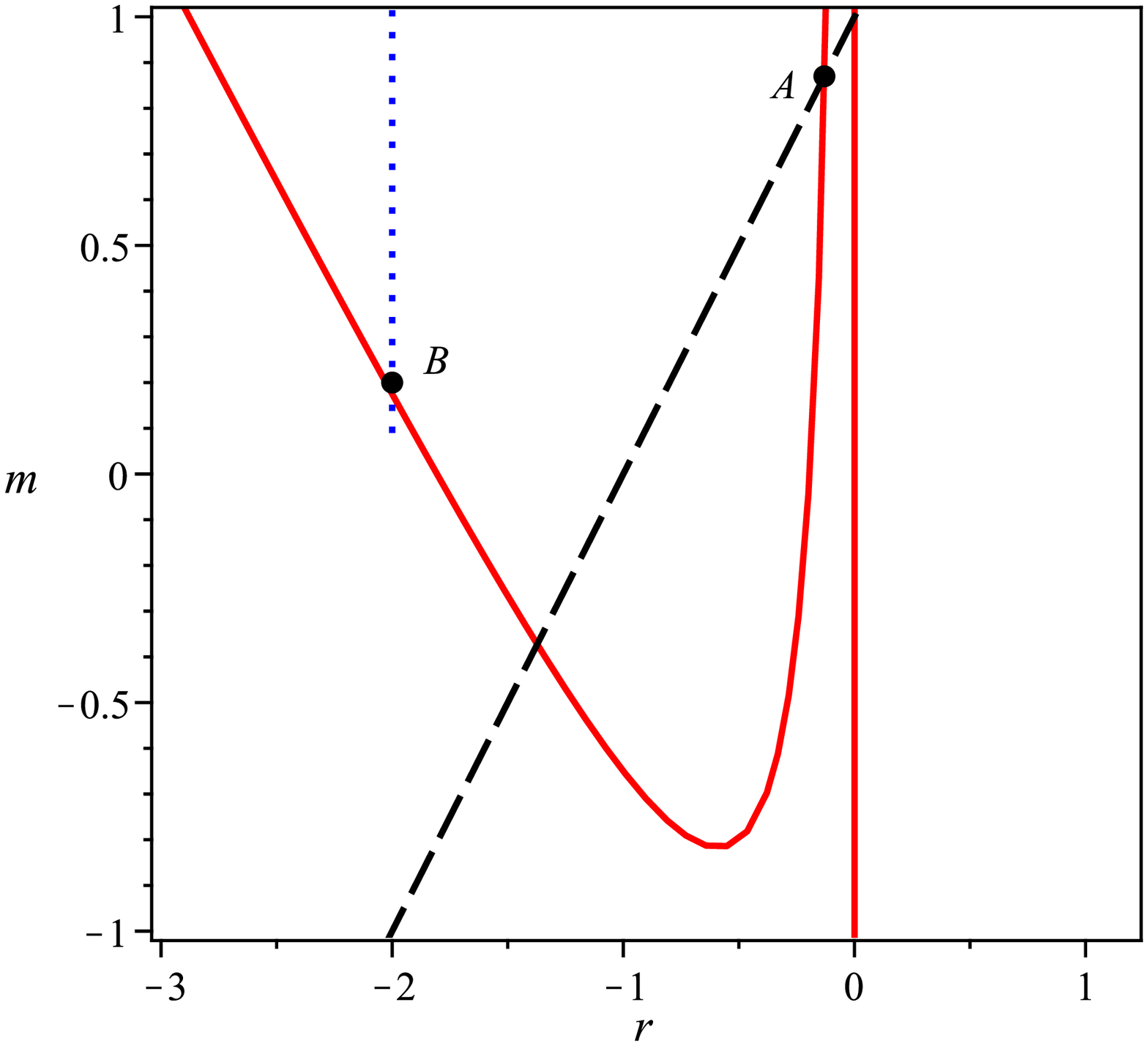}
\vspace{2cm}\includegraphics{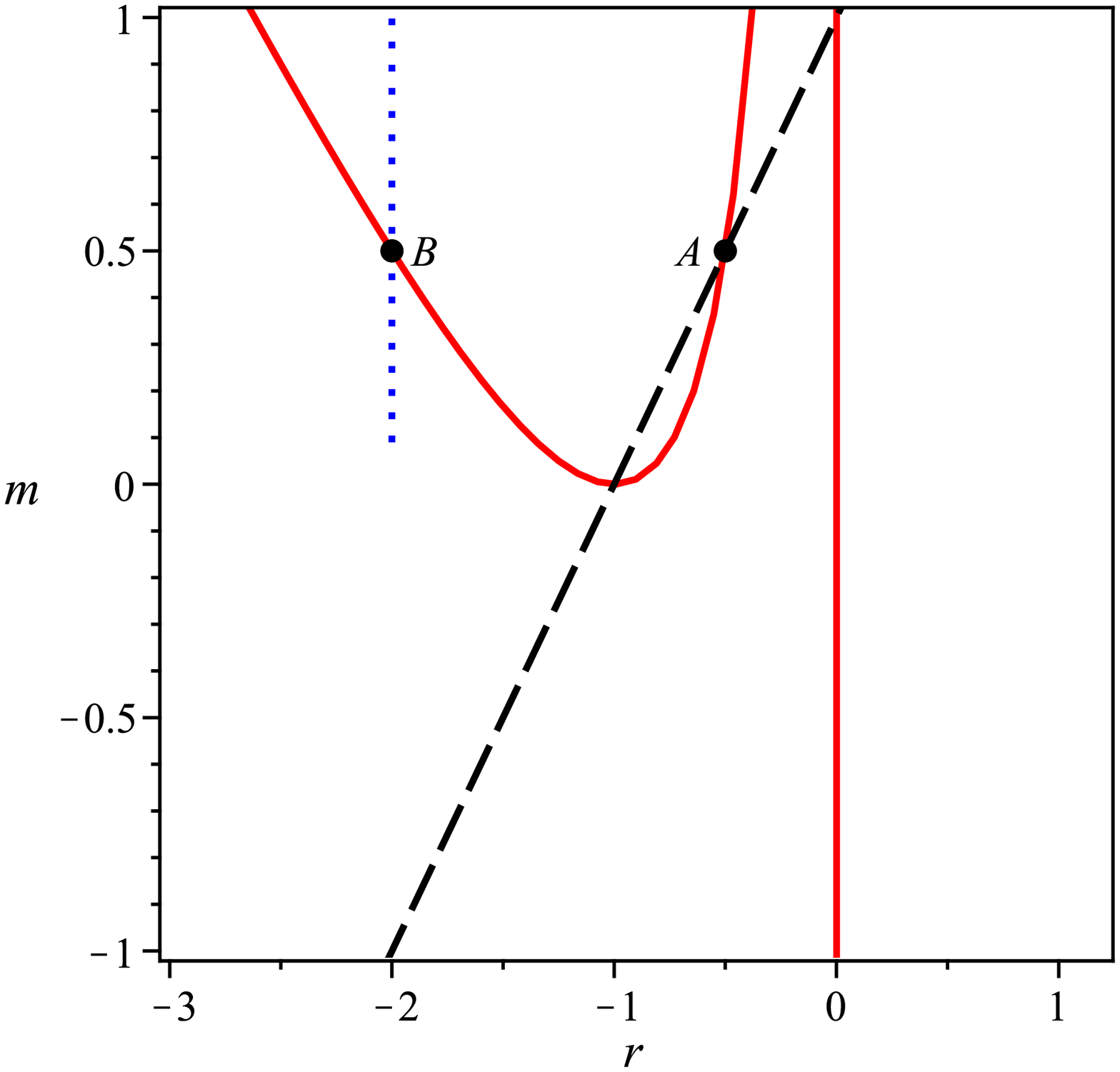} \vspace{3.2cm}
\end{center}
 \caption{\small {The diagram $m-r$ for
$f(R)=R^{n} \exp(\frac{\eta}{R})$ with $n=0.35$ (left panel). The
dashed-line shows the $m(r)$ function of the critical point ${\cal
E}$. The intersection point $A$ shows a matter domination phase.
This model reaches to the standard de Sitter phase at $B$ with
$m=0.2$ which is a stable point. In this model the matter dominated
era can be realized only by ${\cal E}$.}}
\end{figure}
Now, the critical point ${\cal E}$ gives a matter dominated phase
(with $m(r=-0.13)=0.87$) if we set $n=0.35$, that is to say, if
$f(R)=R^{^{0.35}} \exp(\frac{\eta}{R})$\,.  In the model with
$f(R)=R \exp(\frac{\eta}{R})$ (models with
$m(r=-\frac{1}{2})=\frac{1}{2}$), the matter dominated era is
realized by the curve ${\cal D}$ and the critical point ${\cal E}$
plays the role of a non-phantom acceleration era. In figure $5$
(left panel), the curve $m(r)$ is plotted for $f(R)=R^{^{0.35}}
\exp(\frac{\eta}{R})$\,. This model reaches the standard de Sitter
phase at $m=0.2$\,. Therefore, this point is stable since it belongs
to the region defined by relation (27). Finally, we plot the $m(r)$
curve for $f(R)=R \exp(\frac{\eta}{R})$ as shown in the right panel
of figure $5$. In this case there is an acceptable connection
between the matter dominated era and the standard de Sitter phase
since the standard de Sitter point at $m=\frac{1}{2}$ is a stable
attractor. We note also that in this DGP-inspired $f(R)$ model there
is a non-phantom acceleration phase which emerges from the point
${\cal E}$ with $m(r=-\frac{1}{2})=\frac{1}{2}$. This is
corresponding to the point $A$ of figure $5$ (right panel).

\section{Summary and Conclusion}
In this paper we investigated cosmological dynamics of the normal
DGP setup in a phase space approach where the induced gravity is
modified in the spirit of $f(R)$-theories. The motivation for this
study within a dynamical system approach lies in the fact that
recently it has been revealed that the normal, ghost-free DGP branch
has the potential to explain late-time speed-up if we incorporate
possible modification of the induced gravity in the spirit of
$f(R)$-theories. In this respect, a phase space analysis of the
scenario would be interesting to reveal some aspects of this
late-time behavior. Especially the stability of this late-time de
Sitter phase is important to have a cosmologically viable solution.
We applied the dynamical system analysis to achieve the stable
solutions of the scenario in the normal DGP branch. We have shown
that generally there are some fixed points that one of those is the
standard de Sitter phase. Therefore, the \emph{normal branch} of
this DGP-inspired braneworld scenario realizes the late-time
acceleration phase of the universe expansion. However, the stability
of this point depends on the form of $f(R)$ via the parameter
$m\equiv\frac{d \ln{f_{,{R}}}}{d \ln{R}}$. Then, we investigated the
cosmological viability of these setups. A cosmologically viable
scenario contains an early time radiation dominated era followed by
a matter dominated era that reaches a standard de Sitter phase which
is a stable attractor. Here we focused only on the last two stages:
matter domination era followed by a stable de Sitter attractor. To
be more specific, we considered two models with $f(R)=R+\gamma
R^{n}$ and $f(R)=R^{n}\exp (\frac{\eta}{R})$ in our DGP-inspired
setup. The condition for existence of the matter domination era
restricted us to consider two cases $n=0.13$ and $n=\frac{1}{2}$ for
the first model and two cases $n=0.35$ and $n=1$ for the second one.
On the other hand, it is shown that the standard de Sitter phase is
stable just for $m\geq 0.09$. So, since the first model reaches the
standard de Sitter phase (which is determined by the line $r=-2$) at
$m<0.09$\,, it is not a cosmologically viable model. However, since
the second model reaches this phase at $m>0.09$, it is a
cosmologically viable model for $n=0.35$ and $n=1$.

\end{document}